\documentclass[12pt,preprint]{aastex} 
\usepackage{emulateapj5}

\newcommand{\be}{\begin{equation}}
\newcommand{\ee}{\end{equation}}
\newcommand{\w}{{\rm w}}
\def\refnew#1{(\ref{#1})} 
\newcommand{\au}{{\;\rm AU}}
\begin{document}

\title{Planet - Disk Symbiosis} 

\author{Re'em Sari and Peter Goldreich}
\affil{Caltech 130-33, Pasadena, CA 91125}

\begin{abstract}

Planets form in disks around young stars. Interactions with these
disks cause them to migrate and thus affect their final orbital
periods. We suggest that the connection between planets and disks may
be deeper and involve a symbiotic evolution. By contributing to the
outward transport of angular momentum, planets promote disk
accretion. Here we demonstrate that planets sufficiently massive to
open gaps could be the primary agents driving disk accretion. Those
having masses below the gap opening threshold drift inward more
rapidly than the disk material and can only play a minor role in its
accretion. Eccentricity growth during gap formation may involve an
even more intimate symbiosis. Given a small initial eccentricity, just
a fraction of a percent, the orbital eccentricity of a massive planet
may grow rapidly once a mass in excess of the planet's mass has been
repelled to form a gap around the planet's orbit.  Then, as the
planet's radial excursions approach the gap's width, subsequent
eccentricity growth slows so that the planet's orbit continues to be
confined within the gap.

\end{abstract}

\section{Introduction}

One of the most important scientific developments of the nineties was
the discovery of extrasolar planets - planets that orbit other stars.
Currently, more than a hundred are known, along with about a dozen
systems containing more than one planet. The basic data, masses and
orbits, have revealed two major surprises: (I) Jovian mass planets
with short period orbits; (II) Isolated planets with large orbital
eccentricities.

Current planet formation theories fail to account for the formation of
giant planets on short period orbits. Instead, there is a general
consensus that these planets migrated inward from where they were
born. Theoretical work done in the eighties and nineties established
that angular momentum and energy exchanged between planet and disk at
discrete Lindblad resonances would result in the rapid migration of
the planet \citep{GOT80}, and that almost invariably the migration
would be inward \citep{WAR86,ART93a,ART93b}.  In turn, for
sufficiently massive planets, this exchange would modify the disk's
density profile. A familiar aspect of the latter phenomenon is the
opening of a gap around the orbit of a massive planet which locks the
orbital evolution of the planet to that of the disk, a behavior
referred to as type-II migration \citep{WAR97}.

\section{Accretion}

Disk accretion is driven by the outward transport of angular momentum.
The mechanism by which this is accomplished in protostellar disks is
uncertain. Molecular viscosity is far too small to be effective.
Magnetorotational instability is a more plausible candidate, but the
disk's electrical conductivity may be inadequate to sustain it.

Here we examine the possibility that torques from embedded planets
drive disk accretion. This idea is not new. Stimulated by Larson's
\citep{LAR89} suggestion that spiral waves might drive disk accretion,
\cite{GOR01} proposed that these waves could be excited by planets
with masses too small to open gaps. However, they ignored the
migration of the planets. A general argument given below establishes
that subcritical planet's would disappear before significant disk
accretion could take place.

The critical mass required for gap formation in a disk without any
intrinsic viscosity is less than one earth mass
\citep{HOW84,WAH89,RAF02}. Thus subcritical planets are composed of
elements heavier than helium which comprise a fraction $f$ of order a
percent of the disk's mass. Through the torques they exert at Lindblad
resonances, these planets transfer angular momentum outward from
material interior to their orbits to that external to their
orbits. The dominant resonances are located about a distance $h$, the
scale height of the disk, inward and outward from the planet's
orbit. Therefore these two rates are almost equal. Only the small
fractional difference, of order $h/r$, causes the planet to drift
inward. Thus each planet transfers angular momentum across its orbit
at a rate that is larger by a factor $r/h$ than that at which its
angular momentum decays.

Although the major planet-disk interactions occur at distance $h$ from
the planet's orbit, density waves carry the angular momentum farther
away and deposit it at distance $\lambda$ where typically
$h<\lambda<r$.  So the contribution from each planet to the local
luminosity of angular momentum through the disk is a factor
$\lambda/h$ greater than the rate at which it loses angular
momentum. If subcritical planets were the sole source of the disk's
angular momentum luminosity, the timescale for planet accretion would
be related to that for disk accretion by a factor $F\approx
f(\lambda/h)$.  \cite{GOR01} estimate $\lambda/h \sim 2$ for earth
size planets. Since the total mass of subcritical planets is at most a
percent of the disk mass ($f<0.01$), they could not be responsible for
the accretion of more than two percent of the disk mass. 

Although subcritical planets are unable to drive disk accretion,
massive planets might. They reside in gaps and accrete at the same
rate as the surrounding disk material. If they were the sole source of
disk accretion, their number density and masses would affect the
accretion rate but be irrelevant to the final outcome. Suppose there
were one giant planet per logarithmic radius interval. Then gap widths
would be of order the radius and one sided torques from individual
planets would have approximate magnitudes given by $(M_p/M_*)^2 \Sigma
r ^2 (\Omega r)^2$, where $M_p$ and $M_*$ are the masses of the planet
and the star, $\Sigma$ is the disk's surface density, $\Omega$ is the
orbital angular velocity, and $r$ is the orbital radius. Under these
conditions, the angular momentum luminosity in the disk would be of
order the one sided torque. This would yield an accretion timescale of
order $(M_*/M_p)^2 \Omega^{-1}$ provided $M_p \leq \Sigma r^2$. Note
that this timescale is independent of the disk mass and demonstrates
that Jupiter size planets could cause a disk to accrete in about a
million years. Higher rates would occur if the planets were more
massive or numerous subject to the constraint that the mass in planets
does not exceed the disk mass.

\section{Eccentricities and Gaps}

We adopt standard parameters for use in numerical evaluations. They
are: $r=1\au$, $M_d/M_*=10^{-2}$ for disk to star mass ratio where
$M_d$ is the mass inside $r=1\au$, $\mu=M_p/M_*\approx 10^{-3}$ for
the planet to star mass ratio, and $h/r\approx0.04$ for the disk
aspect ratio. Also, in this section, we treat the viscosity with the
customary approximation and choose $\alpha=10^{-4}$ which, combined
with our disk mass and scaleheight, sets a mass accretion rate $\dot M
\approx 10^{-8}$ solar masses per year. This choice is consistent with
observational determinations of accretion rates onto 1My old T Tauri
stars \citep{HCG98}. Higher accretion rates are typical at earlier
stages, but any giant planets that might have formed then would have
been consumed by their parent stars. As we are concerned with planets
that survived, it is the later accretion stages that are relevant to
our investigation . If, as we speculate in the previous section, there
is no intrinsic viscosity and accretion is entirely driven by planet
torques, then the effective $\alpha$ may be even lower than the value
adopted above.

\cite{GoS03} (Hereafter GS) suggest that planet-disk interactions
might have given rise to the large eccentricities of extrasolar
planets. The important interactions are those at Lindblad and
corotation resonances. For sufficiently small eccentricity, the
interactions are linear. In this limit, those at ordinary Lindblad
resonances, which excite eccentricity, are slightly less effective
than those at corotation resonances, which damp it. However, nonlinear
saturation of the corotation resonances occurs at small eccentricities
(\cite{GoT81,OgL03}; GS) so a finite amplitude instability leading to
eccentricity growth is a distinct possibility. Perhaps the most
serious concern with the scenario by which eccentricity grows due to
planet disk interactions involves the relative importance of
eccentricity damping due to apsidal waves, although it is weaker than
previously thought (GS).

GS consider eccentricity evolution in the context of steady-state
gaps. Here we argue that conditions during gap formation are more
favorable for eccentricity growth. This is significant since it is the
initial stage of eccentricity growth that is the most problematic.

Next we estimate the minimal eccentricity required to saturate the
corotation resonances when a gap has just formed. At this stage its
width, $\w \sim a/m \ll r$, is the larger of the disk's scale height
and the planet's Hill radius, each of which are much smaller than the
equilibrium width. Corotation saturation occurs when the density
gradient is flattened over a scale $\delta=(em\mu)^{1/2}r$. Our story
involves the comparison of several rates, or inverse timescales,
namely:

(i) The rate at which the density gradient is flattened at a
first order corotation resonance by the corotation torque,
\begin{equation}
t_{sat}^{-1} \approx 
 \mu^{1/2}e^{1/2}\left( r \over \w \right)^{3/2} \Omega.
\label{eq:tsat}
\end{equation}

(ii) The rate at which viscosity reestablishes the original density
gradient over a scale $\delta$,
\begin{equation}
t^{-1}_{vis}={\nu \over \delta^2}\approx {\alpha\over \mu
e}\left({\w\over r}\right) \left( h \over r\right)^2 \Omega.
\label{eq:tvis}
\end{equation}

(iii) The rate at which a gap of size $\w$ is opened by principal
Lindblad resonances,
\begin{equation}
t_{gap}^{-1} \equiv {1 \over \w} {d\w \over dt} 
\approx  \mu^2 \left( r\over \w \right)^5 \Omega
\label{eq:tgap}
\end{equation}

(iv) The eccentricity growth rate assuming only Lindblad resonances
are active,
\begin{equation}
\label{eq:te}
t_e^{-1} \equiv {1 \over e}{de \over dt} \approx 
\mu^2\left( r \over \w \right)^4 \left(M_d \over M_p \right) \Omega.
\end{equation}
When corotation resonances are unsaturated, eccentricity damps at a rate
which is smaller by a factor of about twenty \citep{GOT80}.

For the eccentricity to grow, the corotation resonances must
saturate. During gap formation, this requires both $t_{sat}<t_{vis}$
and $t_{sat}<t_{gap}$. Only the former is relevant for a steady state
gap, and it leads to the criterion derived by \cite{GoT81} and
\cite{OgL03}. GS estimate a critical initial eccentricity of about
$1\%$ for eccentricity growth in equilibrium gaps where torques at
principal Lindblad resonances balance the large scale viscous
torque. However, as we now show, smaller gaps lead to less stringent
constraints on the required initial eccentricity.

The requirement $t_{sat}<t_{vis}$ is satisfied provided
\begin{equation}
\label{tsatvis}
e>{\alpha^{2/3}\over\mu}\left(\w\over r\right)^{5/3}\left(h\over
r\right)^{4/3},
\end{equation}
and $t_{sat}<t_{gap}$ provided
\begin{equation}
\label{tsatgap}
e>\mu^3\left(r\over\w\right)^7.
\end{equation}
Thus saturation is most readily achieved at a width 
\begin{equation}
\label{wbest}
{\w\over r}\sim {\mu^{6/13}\over \alpha^{1/13}}\left(r\over
h\right)^{2/13}\sim 0.14, 
\end{equation}
provided the initial eccentricity satisfies
\begin{equation}
e_0>{\alpha^{7/13}\over\mu^{3/13}}\left(h\over r\right)^{14/13}\sim 10^{-3}.
\label{ecrit1}
\end{equation}

Saturation of corotation resonances is not a sufficient condition for
significant eccentricity growth during gap formation. In addition, the
growth rate of eccentricity must exceed that of the gap. This occurs
for
\begin{equation}
\label{wmp}
{\w \over r} >{M_p \over M_d } \sim 0.1 ,
\end{equation} 
in other words, once the mass expelled from the gap exceeds the
planet's mass. Corotation resonances are saturated at this width
provided the initial eccentricity satisfies
\begin{equation}
e_0>{\alpha^{2/3}\over \mu}\left(h\over r\right)^{4/3}\left(M_p\over
M_d\right)^{5/3}\sim 6 \times 10^{-4}.
\label{ecrit2}
\end{equation}
Significant eccentricity growth requires the initial eccentricity to
satisfy both inequalities \refnew{ecrit1} and \refnew{ecrit2}.

Note that eccentricity decays if corotation resonances are less than
$5\%$ saturated. However, significant decay during gap growth can only
occur once $\w>20(M_p/M_d)r$. This is avoided provided the initial
eccentricity satisfies the additional constraint
\begin{equation}
e_0> 10^{-12} \mu^{-4} \left( M_d \over M_\odot \right)^7 \sim 10^{-14},
\end{equation}
which is relevant only for lighter planets or more massive disks (e.g.
a planet with mass ten times smaller than Jupiter's in a tenth of a
solar mass disk). Subsequently, we ignore the possibility of
significant eccentricity decay during gap formation.

How does a planet's eccentricity grow as its gap widens after the corotation
resonances are fully saturated?  By dividing the gap growth rate by the
eccentricity growth rate we arrive at
\begin{equation}
e=e_0  \exp\left({\w-\w_0 \over a} {M_d \over M_p} \right).
\label{efold}
\end{equation}
If the disk is ten times more massive than the planet, by the time the
gap approaches its equilibrium width, comparable to the radius of the
disk, equation \refnew{efold} predicts that the eccentricity would
have grown by of order ten e-folds. The resulting radial excursions of
the planet would cause it to pass beyond the gap edges. We expect that
before this happens, additional damping that we have not modeled would
come into play and limit eccentricity growth. From then on it is
likely that the eccentricity would be coupled to the gap width such
that $e\sim \w/r$. Our conclusions about eccentricity evolution as a
function of gap width are depicted in figure 1.

\begin{figure*}
\begin{center}
\epsscale{1}
\plotone{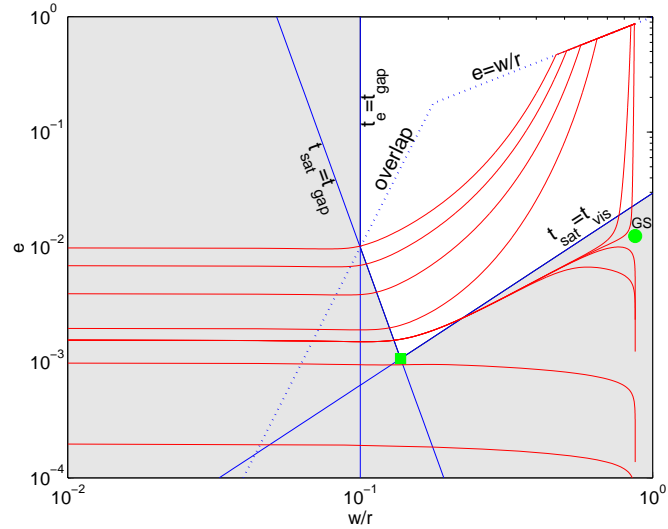}
\end{center}
\caption{Evolution of a planet's eccentricity versus fractional gap
width for a variety of initial values of eccentricity. Fast
eccentricity growth occurs to the right of the vertical line, where
the mass expelled from the gap exceeds the planet's mass, and above
the two solid diagonal lines (equations \refnew{tsatvis} and
\refnew{tsatgap}), where the corotation resonances are fully
saturated. In that region of the $(e,\w)$ plane, shown as unshaded,
eccentricity grows exponentially until it approaches the fractional
gap width shown by the dotted line marked $e=\w/r$.
The green circle marks the criterion for
eccentricity growth derived by GS for steady-state gaps. However,
since significant eccentricity growth can occur before the gap reaches
its maximum width, the actual criterion is less stringent, and is
indicated by the green square. The criterion for resonance overlap is
shown as dotted line marked ``overlap''.}
\label{fig:gap}
\end{figure*}

To conclude the discussion of corotation resonances, we mention the
possibility of resonance overlap. This seems to be particularly
relevant here as we are interested in eccentricity evolution when the
gaps are narrow, the fractional distance between resonances,
$\w^2/r^2$, is small, and the fractional width of the resonances, 
$(\mu e r/\w)^{1/2}$, is relatively
large. Resonances overlap if
\begin{equation}
e> {1\over \mu} \left(\w\over r\right)^5.
\end{equation}
The evolution of eccentricity in this regime requires further
investigation. However, as shown in the figure, for our fiducial
parameters most of the evolution occurs below this eccentricity, so
resonance overlap probably does not play an important role even during
gap formation.

Apsidal waves also promote eccentricity damping. GS found their
influence to be marginal for a gap at its equilibrium width. Since the
relative importance of apsidal waves varies as the cube of the gap
width, they are unlikely to play a major role when the gap is well
below its equilibrium size. Of course, for a planet to maintain its
orbital eccentricity, the apsidal waves must also be unimportant at
the equilibrium gap width. But provided the eccentricity can grow when
the gap is narrow, by the time the gap approaches its full width the
corotation resonances will be fully saturated. Then eccentricity
maintenance merely requires that eccentricity damping by apsidal waves
be weaker than eccentricity excitation by first order Lindblad
resonances, a criterion that is satisfied provided (GS):
\begin{equation}
\left( r \over h\right)^4 \left( M_p \over M_* \right)^2 < 12.
\end{equation}
This requirement is about a factor of twenty less stringent than that
derived by GS for the apsidal wave torque to be less important than
the small difference between Lindblad and partially saturated
corotation resonance torques. In particular, this looser criterion is
amply satisfied by Jupiter mass planets at $r=1\au$.

\section{Planet Disk Symbiosis}

We have shown that a symbiotic relation might exist between massive
planets and the disks in which they form. This relation would explain
the similarity between timescales estimated to be needed for planet
growth and those derived from observations of disk accretion, because
disk accretion would only commence after massive planets form. In this
scenario, most planets commit suicide by promoting the accretion of
the disk to which they are locked. That some planets survive implies
that not all of the disk material is accreted. Some other mechanism,
such as evaporation, might remove the final remnants of the disks.

The interplay between eccentricity excitation and gap formation is a
more subtle aspect of this symbiosis. Once a planet attains a
sufficient mass, perhaps a few earth masses, it rapidly accretes an
envelope of hydrogen and helium and begins to open a gap in its
disk. Provided the planet's orbit is endowed with a small initial
eccentricity, the corotation resonances saturate while the gap width
is not much larger than the disk scale height $h$. However, despite
their saturation, significant eccentricity growth does not occur until
the mass cleared from the gap becomes comparable to the planet's
mass. Subsequently, the rate of fractional eccentricity increase
exceeds that of the gap width. From then on it is plausible that the
eccentricity maintains a value of order the fractional gap width,
$e\approx \w/r$. A schematic description of eccentricity and gap
growth is displayed in the figure.

Eccentricity growth during gap opening alleviates two of the major
concerns raised by \cite{GoS03} in their discussion of possible
eccentricity growth for planets in steady-state gaps. It reduces both
the relative importance of the apsidal torque and the required value
of the initial eccentricity. The case for eccentricity growth due to
planet disk interactions becomes more promising with this new
analysis.

\acknowledgements
This research was supported in part by an NSF grant AST-0098301 and
NASA grant NAG5-12037.

\bibliographystyle{apj}
\bibliography{planets}

\end{document}